\documentclass[twocolumn,showpacs,prl,aps,superscriptaddress]{revtex4-1}
\usepackage{amsmath}
\usepackage{amsfonts}
\usepackage{amssymb}
\usepackage{graphicx}
\usepackage{hyperref}
\usepackage{bm}
\usepackage{color}
\usepackage{times}

\hypersetup{backref,
colorlinks=true,
linkcolor=blue,
linktoc=black,
citecolor=cyan,
urlcolor=black}

\DeclareMathOperator{\sinc}{sinc}

\newcommand{\pr}[1]{\ensuremath{\left[#1\right]}} 
\newcommand{\pc}[1]{\ensuremath{\left(#1\right)}}

\newcommand{\ket}[1]{\ensuremath{\left\vert#1\right\rangle}} 
\newcommand{\md}[1]{\ensuremath{\left\vert#1\right\vert}}

\begin{document}
\title{
Quantum depletion of a homogeneous Bose--Einstein condensate
}

\author{Raphael Lopes}
\email{rl531@cam.ac.uk}
\affiliation{Cavendish Laboratory, University of Cambridge, J. J. Thomson Avenue, Cambridge CB3 0HE, United Kingdom }
\author{Christoph Eigen}
\affiliation{Cavendish Laboratory, University of Cambridge, J. J. Thomson Avenue, Cambridge CB3 0HE, United Kingdom }
\author{Nir Navon}
\affiliation{Cavendish Laboratory, University of Cambridge, J. J. Thomson Avenue, Cambridge CB3 0HE, United Kingdom }
\affiliation{Department of Physics, Yale University, New Haven, CT 06511, USA}
\author{David Cl\'ement}
\affiliation{Laboratoire Charles Fabry, Institut d'Optique Graduate School, CNRS, Universit\'e Paris-Saclay, 91127 Palaiseau cedex, France }
\author{Robert P. Smith}
\affiliation{Cavendish Laboratory, University of Cambridge, J. J. Thomson Avenue, Cambridge CB3 0HE, United Kingdom }
\author{Zoran Hadzibabic}
\affiliation{Cavendish Laboratory, University of Cambridge, J. J. Thomson Avenue, Cambridge CB3 0HE, United Kingdom }

\begin{abstract}
We have measured the quantum depletion of an interacting homogeneous Bose--Einstein condensate, and confirmed the 70-year old theory of N.~N. Bogoliubov. The observed condensate depletion is reversibly tuneable by changing the strength of the interparticle interactions.
Our atomic homogeneous condensate is produced in an optical-box trap, the interactions are tuned via a magnetic Feshbach resonance, and the condensed fraction probed by coherent two-photon Bragg scattering. 
\end{abstract}
\maketitle

After superfluidity of liquid $^4$He was discovered in 1937~\cite{Kapitza:1938,Allen:1938}, its connection to Bose--Einstein condensation was posited by F.~London~\cite{London:1938} and L.~Tisza~\cite{Tisza:1947}. However, while at zero temperature liquid helium is $100\%$ superfluid, less than $10\%$ of the atoms are actually in the Bose--Einstein condensate (BEC)~\cite{Miller:1962}; most of the particles are coherently expelled from the condensate by strong interactions, and spread over a wide range of momenta.
In 1947 N.~N.~Bogoliubov developed a theory that explains the microscopic origin of such interaction-driven, quantum depletion of a BEC~\cite{Bogoliubov:1947}. 
This theory has become a cornerstone of our conceptual understanding of quantum fluids, but is quantitatively valid only for relatively weak interactions, and could not be tested with liquid helium.

Nowadays, gaseous atomic BECs provide a flexible setting for exploring the rich physics of interacting Bose fluids~\cite{Dalfovo:1999,Bloch:2008, Chevy:2016}, and many liquid-helium-inspired  theories can now be directly confronted with experiments. According to the Bogoliubov theory, for a homogeneous Bose gas of particle density $n$ and interactions characterised by the scattering length $a$, and assuming $\sqrt{na^3} \ll 1$, the condensed fraction at zero temperature is~\cite{Lee:1957b}
\begin{equation}
n_\text{BEC}/n = 1 - \gamma \sqrt{na^3}\, ,
\label{eq:Bog}
\end{equation}
where $\gamma = 8/(3\sqrt{\pi}) \approx 1.5$. 
Effects of quantum depletion have been observed in harmonically trapped atomic gases, both by enhancing the role of interactions in optical lattices~\cite{Xu:2006} (see also~\cite{Greiner:2002a}) and in high-resolution studies of the expansion of a weakly-interacting gas~\cite{Chang:2016}.  However, only semiquantitative comparison with theory has been possible, due to complications associated with the addition of the lattice, the inhomogeneity of the clouds, and/or the interpretation of the expansion measurements~\cite{Qu:2016}.

In this Letter, we test and verify the Bogoliubov theory of quantum depletion in a textbook setting, using a homogeneous  $^{39}$K BEC~\cite{Eigen:2016}, produced in an optical-box trap~\cite{Gaunt:2013}, and tuning the interaction strength via a magnetic Feshbach resonance~\cite{Chin:2010}. To measure the condensed fraction of our clouds we use `BEC filtering'~\cite{Gerbier:2004c} - using Doppler-sensitive two-photon Bragg scattering~\cite{Kozuma:1999a,Stenger:1999b} we {\it spatially} separate the BEC from the high-momentum components of the gas.

  \begin{figure}[t!]
\centering 
  \includegraphics[width=\columnwidth]{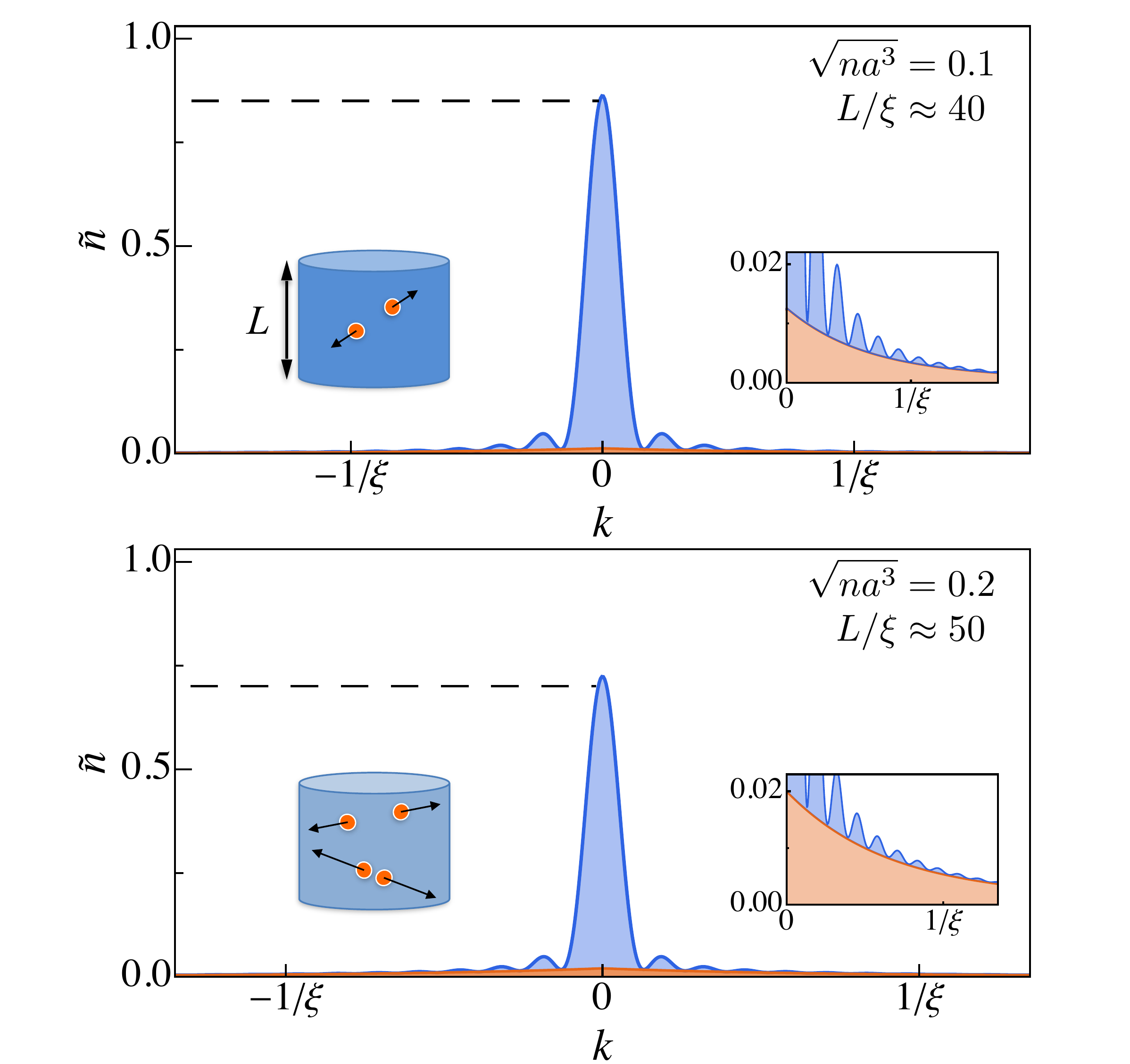}
\caption{\label{Fig1}
Momentum distribution of a zero-temperature homogeneous Bose gas. We consider a gas of density $n$ and size $L$, and two different values of the scattering length $a$. We show the expected 1D momentum distribution $\tilde{n}(k)$ (see text), normalised so that $\tilde{n}(0) = 1$ would correspond to no quantum depletion (setting $\gamma$ in Eq.~(\ref{eq:Bog}) to 0). The total $\tilde{n}(k)$ consists of the BEC peak (blue), with a Heisenberg-limited width $\propto 1/L$, and a broad quantum-depletion pedestal (orange) of characteristic width $1/\xi$, where $\xi$ is the healing length. To a good approximation, the low-$k$ distribution is the same as for a pure BEC, just scaled by a factor $1-\gamma\sqrt{na^3}$, indicated by the dashed lines.
For this illustration we use experimentally relevant values of $L/\xi$, but exaggerated values of $\sqrt{na^3}$, to make the orange shading visible in the main panels. Also note that we assume that the very broad $\tilde{n}_{\rm QD}(k)$ is not affected by finite-size effects.
The cartoons on the left depict the coherent excitations out of the (blue) condensate, which occur as pairs of atoms with opposite momenta.
The right insets highlight the fact that $\tilde{n}_{\rm QD}(k) \gg \tilde{n}_{\rm BEC}(k)$ at large $k$.
}
 \end{figure}

We produce our homogeneous clouds in a cylindrical box trap (see Fig.~\ref{Fig1}) of radius $R = 32~\mu$m and length $L=50~\mu$m~\cite{Eigen:2016}, and probe them using Bragg scattering along the axis of the cylinder ($z$)~\cite{Lopes:2017}. Bragg diffraction imparts to an atom a momentum $\hbar q$, where
(in our setup) $q= 1.7 \times 2\pi / \lambda$ and $\lambda = 767$~nm. 
The frequency difference between the two Bragg beams selects the initial momentum of an atom for which the scattering is efficient~\cite{Kozuma:1999a,Stenger:1999b}, with momentum resolution $\Omega m/q$, where $\Omega$ is the two-photon Rabi frequency and $m$ the atom mass. 
The Bragg resonance condition depends only on an atom's initial momentum along $z$, so we effectively probe the one-dimensional (1D) momentum distribution of the cloud, $\tilde{n}(k)$, given by the integral of the 3D distribution along the two transverse directions.

\begin{figure}[t!]
\centering
\includegraphics[width=1\columnwidth]{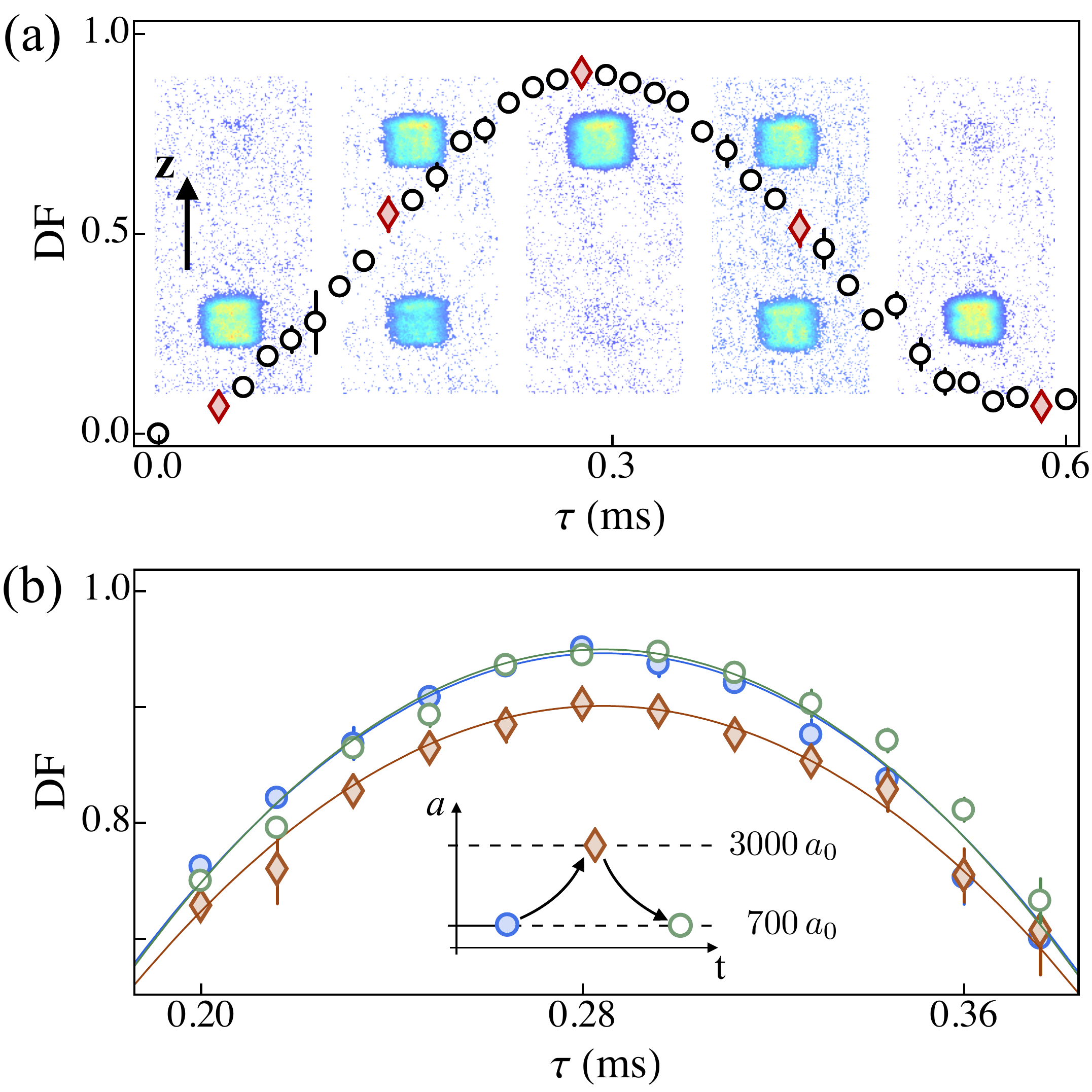}
\caption{
Bragg filtering and reversible interaction-tuning of the condensed fraction. (a) Diffracted fraction (DF) as a function of the Bragg pulse duration, $\tau$, for $\Omega = 2\pi \times 1.8$~kHz and $a \approx 3000~a_0$. 
Absorption images in the background 
show the stationary (bottom) and diffracted (top) clouds, for the data points indicated by the red diamonds.
(b) Diffracted fraction for $\tau$ close to $\pi/\Omega$, for three different preparations of the cloud (see inset): at $700~a_0$ (filled blue circles), after raising $a$ from $700~a_0$ to $3000~a_0$ in 80~ms (orange diamonds), and after reducing it back to $700~a_0$ in another $80$~ms (open green circles). We see that increasing $a$ reversibly reduces the maximal  diffracted fraction. All error bars show standard statistical errors in the mean.}
\label{Fig2}
\end{figure}

Spatially separating the BEC from the quantum depletion (QD) relies on a separation of three momentum scales, $1/L \ll 1/\xi \ll q$, where $\xi = 1/\sqrt{8\pi na}$ is the healing length.  In Fig.~\ref{Fig1} we illustrate the expected $\tilde{n}(k)$ for a zero-temperature gas: 
$\tilde{n}(k) = \tilde{n}_{\rm BEC}(k) + \tilde{n}_{\rm QD}(k)$, where $\tilde{n}_{\rm BEC}$ has a Heisenberg-limited width $\propto 1/L$~\cite{Gotlibovych:2014} and exponentially suppressed high-$k$ tails, while $\tilde{n}_{\rm QD}(k)$ has a width $\propto 1/\xi$ and long polynomial tails~\cite{Pethick:2002,Wild:2012,Makotyn:2014,Chang:2016} (see~\cite{Lopes:2017QD_SI} for details). 
The inequality $L/\xi \gg 1$ thus ensures that $\tilde{n}_{\rm QD}(k)$ extends over a much wider range of momenta than  $\tilde{n}_{\rm BEC}(k)$, so $\Omega$ can be chosen such that a Bragg pulse tuned in resonance with $k=0$ diffracts essentially the whole BEC and almost none of the QD.
The inequality $q\xi \gg 1$ ensures that the momentum kick received by a diffracted atom, $\hbar q$, is much larger than the QD momentum spread, so after the Bragg pulse and subsequent time-of-flight the diffracted and the non-diffracted portions of the cloud can clearly separate in real space [see Fig.~\ref{Fig2}(a)].
For all our measurements  $L/\xi > 30$ and $q\xi > 12$.

We start by producing a quasi-pure weakly-interacting BEC of density $n\approx 3.5 \times 10^{11}$~cm$^{-3}$ in the lowest $^{39}$K hyperfine state, $\ket{F = 1, m_F =1}$ in the low-field basis, which features a Feshbach resonance centred at $402.70(3)$~G~\cite{Fletcher:2017}. 
We prepare the BEC at $a= 200~a_0$, where $a_0$ is the Bohr radius, so $\sqrt{na^3} < 10^{-3}$, and in time-of-flight expansion we do not discern any thermal fraction.
We then (in $150-250$~ms) increase $a$ to a value in the range $700 - 3000~a_0$, and measure the condensed fraction. Our largest $a$ is limited by imposing requirements that: (i) during the whole experiment the atom loss due to three-body recombination is $<10\%$, and (ii) if we reduce $a$ back to $200~a_0$ we do not observe any signs of heating.
To prepare the initial quasi-pure BEC we lower the trap depth $U_0$ to $\approx k_{\rm B} \times 20$~nK, but before increasing $a$ we adiabatically raise $U_0$ by a factor of 5, to ensure that $U_0 \gg \hbar^2/(2m\xi^2)$.

Just before turning off the trap and applying the Bragg pulse, we rapidly (in $60~\mu$s) turn off the interactions, using a radio-frequency pulse to transfer the atoms to the $\ket{F=1,\,m_F=0}$ state, in which $a\approx 0$~\cite{Fletcher:2017}. This freezes the momentum distribution before we probe it by Bragg diffraction, and ensures that the diffracted and non-diffracted components of the gas separate in space without collisions. 

After the Bragg pulse, we wait for $10$~ms and then take an absorption image along a direction perpendicular to $z$ [see Fig.~\ref{Fig2}(a)]. In $10$~ms the diffracted and non-diffracted portions of the gas separate by $\approx 220~\mu$m, while neither expands significantly beyond the original size of the box-trapped cloud.


In Fig.~\ref{Fig2}(a) we show a typical variation of the diffracted fraction of the gas with the duration of the Bragg pulse, $\tau$, for our chosen $\Omega = 2 \pi \times 1.8$~kHz (see also~\cite{Lopes:2017QD_SI}). In the background we show representative absorption images of the stationary (bottom) and diffracted (top) clouds.


Assuming that we perfectly filter out the condensate from the high-$k$ components of the gas, the condensed fraction of the cloud is given by the maximal diffracted fraction, $\eta$, observed for $\tau = \pi/\Omega \approx 0.28$~ms. We see that $\eta$ is slightly below unity, which is expected due to quantum depletion, but can in practice also be observed for other reasons, including experimental imperfections and the inevitably nonzero temperature of the cloud. It is therefore important that our measurements are differential - we study the variation of $\eta$ with $a$, while keeping other experimental parameters the same. It is also crucial to verify that the tuning of $\eta$ with $a$ is adiabatically reversible, which excludes the possibility that the condensed fraction is reduced due to non-adiabatic heating or losses.

In Fig.~\ref{Fig2}(b) we focus on $\tau \approx \pi/\Omega$, and show measurements for three different experimental protocols: for a cloud prepared at $700~a_0$, after increasing $a$ to $3000~a_0$, and after reducing it back to $700~a_0$ (see inset). We see that $\eta$ is indeed reduced when $a$ is increased, and also that this effect is fully reversible (within experimental errors); we have verified such reversibility for our whole experimental range of $a$ values.

 \begin{figure}[t!]
\centering
  \includegraphics[width=\columnwidth]{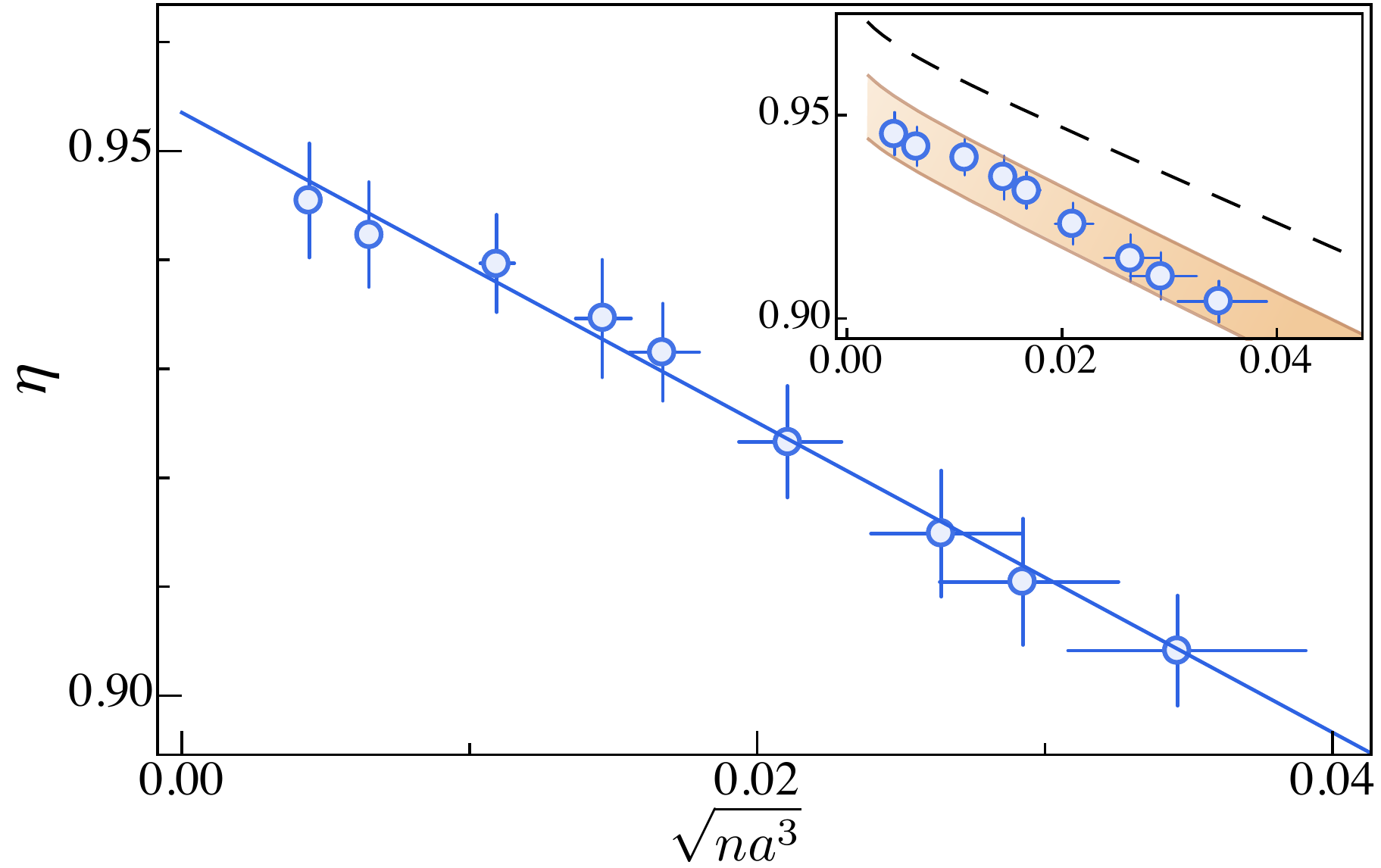}
  \caption{
Measurement of the quantum depletion.
We plot the maximal diffracted fraction $\eta$ versus the interaction parameter $\sqrt{na^3}$. A linear fit (solid line) gives $\eta(0) = 0.954(5)$  and $\gamma = 1.5(2)$. 
Vertical error bars show fitting errors, while horizontal ones reflect the uncertainty in the position of the Feshbach resonance and a $10\%$ uncertainty in $n$. Inset: Analysis of systematic effects. We show numerical simulations for $T=0$ (dashed line) and for initial temperatures (at $a=200~a_0$) between 3.5 and 5~nK (orange shading, from top to bottom); see text and~\cite{Lopes:2017QD_SI} for more details.}
\label{Fig3}
\end{figure}

In Fig.~\ref{Fig3} we summarise our measurements of the variation of $\eta$ with the interaction parameter $\sqrt{na^3}$.
We observe the expected linear dependence, with $\eta(0)$ close to unity. Fitting the data with $\eta(0) ( 1 - \gamma \sqrt{na^3})$ gives $\gamma = 1.5(2)$, in agreement with Eq.~(\ref{eq:Bog}).

Finally, we numerically assess the systematic effects on $\gamma$ due to non-infinite $L/\xi$ and a small nonzero temperature  $T$, which are both $\lesssim 20\%$, and partially cancel. 
The results of this analysis are shown in the inset of Fig.~\ref{Fig3}; for details see~\cite{Lopes:2017QD_SI}. 
The dashed line shows the simulated $\eta$ for $T=0$ and our values of $n$, $L$ and $\Omega$. For any non-infinite $\Omega$, the tails of the BEC momentum distribution are not fully captured by the Bragg pulse, which slightly reduces $\eta (0)$. More importantly, we diffract some of the quantum-depletion atoms, which reduces the apparent $\gamma$. A linear fit (omitted for clarity) gives that for $T=0$ we actually expect $\gamma \approx 1.2$. 
The small systematic differences between our data and this simulation can be explained by a small nonzero temperature. 
A nonzero temperature generally reduces $\eta$ due to thermal depletion, the momentum tails of which are not diffracted by the Bragg pulse. Moreover, if the gas is initially prepared (at $200~a_0$) at a small $T>0$, this does not merely reduce $\eta$ by a constant offset (independent of $\sqrt{na^3}$), but slightly increases the apparent $\gamma$; even adiabatically increasing $a$ increases the thermal depletion, because it modifies both the dispersion relation and the particle content of the thermally populated low-$k$ excitations~\cite{Pethick:2002, Lopes:2017QD_SI}.
As indicated by the orange shaded region, our data are consistent with an initial $T$ between 3.5 and 5~nK; this is compatible with the fact that we do not discern the corresponding thermal fractions of $\lesssim 10\%$ in time-of-flight expansion at $200~a_0$, and is reasonable for our trap depth of $\approx 20$~nK.
Due to these effects the expected dependence of $\eta$ on $\sqrt{na^3}$ is also not perfectly linear, but this effect is negligible on the scale of the experimental errors.

In conclusion, within a $15\%$ statistical error and $20\%$ systematic effects, we have quantitatively confirmed the Bogoliubov theory of quantum depletion of a Bose--Einstein condensate, which is one of the cornerstones of our understanding of interacting quantum fluids. The methods employed here could be extended to study the momentum distribution of the quantum depletion, and could also be useful for sensitive thermometry of homogeneous ultracold Bose gases.

We thank Richard Fletcher and Fabrice Gerbier for inspiring discussions. This work was supported by the Royal Society, EPSRC [Grant No. EP/N011759/1], ERC (QBox), AFOSR, and ARO. R.L. acknowledges support from the E.U. Marie-Curie program [Grant No. MSCA-IF-2015 704832] and Churchill College, Cambridge. N.N. acknowledges support from Trinity College, Cambridge. D.C. acknowledges support from the Institut Universitaire de France.


%


\newpage
\cleardoublepage

\setcounter{figure}{0} 
\setcounter{equation}{0} 

\renewcommand\theequation{S\arabic{equation}} 
\renewcommand\thefigure{S\arabic{figure}} 

\section{Supplemental Information}

\subsection{BEC and quantum-depletion momentum distributions}

Assuming a top-hat BEC wavefunction $\psi(z)$, of extension $L$, gives $\tilde{n}_\text{BEC} (k) \propto  \sinc^2 \pc{kL/2}$. This is a good approximation at low $k$, but the unphysical sharp edges in real space give unphysical high-$k$ tails, $\propto 1/k^2$. In reality the wavefunction is rounded-off near the trap walls and the high-$k$ tails are exponentially suppressed. To take this into account, we write $\psi (z) \propto \tanh \pc{ \frac{L/2 - \md{z}}{\sqrt{2}\xi}}$~\cite{Pethick:2002}, for $\md{z} < L/2$, and numerically compute $\tilde{n}_\text{BEC} (k)$.

Since quantum depletion (QD) is spread over a very wide range of momenta, finite-size effects are negligible, and we use the standard textbook prediction~\cite{Pethick:2002} for the three-dimensional QD distribution in an infinite system; after integration over two directions this gives
\begin{equation}
\tilde{n}_\text{QD} (k) \propto a \pc{1+ \xi^2 k^2  - \sqrt{\xi^2 k^2 \pc{\xi^2 k^2 +2}}}  .
\label{eq:nQD}
\end{equation}

The expression on the right hand side of Eq.~(\ref{eq:nQD}) is normalised to $\gamma \sqrt{na^3}$, with $\gamma \approx 1.5$, and we normalise the numerical BEC momentum distribution to $1- \gamma \sqrt{na^3}$. For intuitive visual presentation, in Fig.~\ref{Fig1} of the main paper and Fig.~\ref{Fig1S} we scale both distributions so that in absence of quantum depletion $\tilde{n}(0) = 1$.


\subsection{Bragg-diffraction efficiency and choice of $\Omega$}

For a Bragg pulse tuned in resonance with the $k=0$ atoms, the Doppler detuning for $k\neq 0$ is $\hbar k q/m$, so for the pulse duration $\tau = \pi/\Omega$ (an on-resonance Rabi $\pi$-pulse), the $k$-dependent diffracted fraction is
\begin{equation} \nonumber
p_\text{Bragg} (k) = \frac{\Omega^2}{\Omega^2 + \pc{\hbar q k/m}^2} \sin^2 \pr{ \sqrt{\Omega^2 + \pc{\hbar q k/m}^2}  \frac{\pi}{2\Omega}} .
\end{equation}

\begin{figure}[t!]
\centering
  \includegraphics[width=\columnwidth]{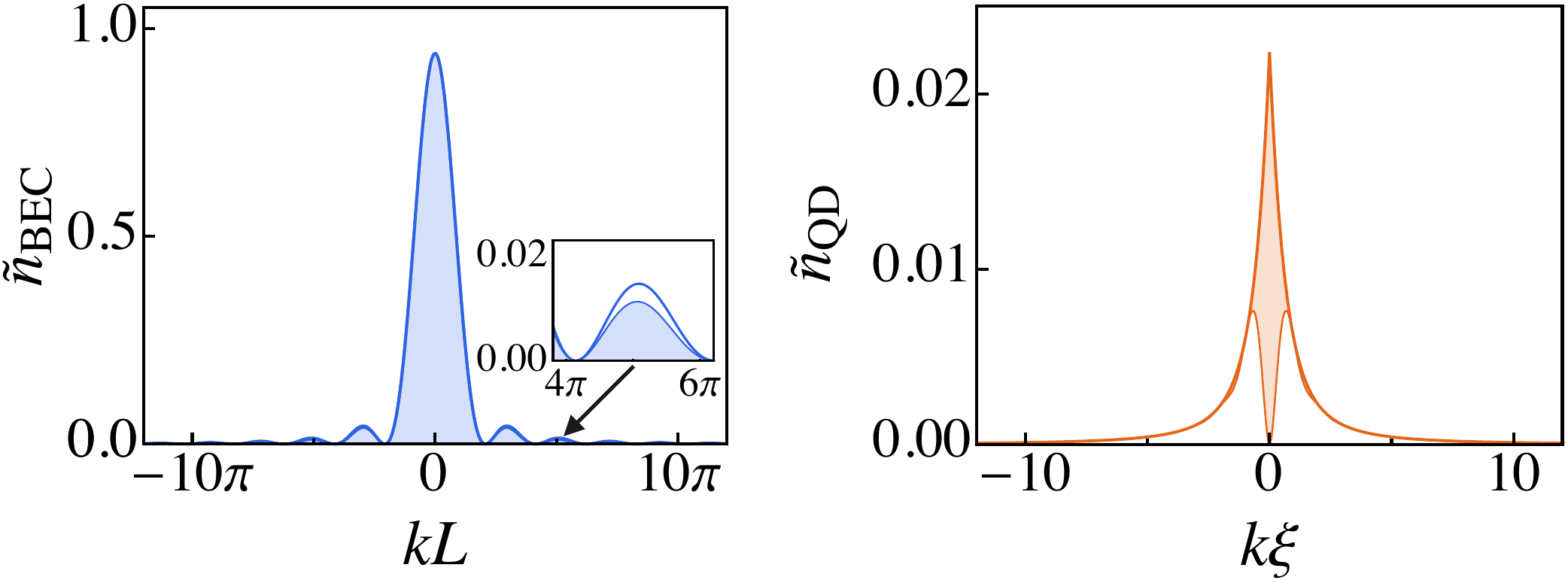}
\caption{\label{Fig1S}
Momentum distributions and diffraction efficiency. Zero-temperature 1D momentum distributions of the BEC (left) and the QD (right), for 
 $n=3.5 \times 10^{11} \text{cm}^{-3}$, $L=50~\mu$m, and $\sqrt{na^3} = 0.04$ (solid lines). Note the difference in the $x$-axes. The shadings indicate the fractions of the BEC and QD that are diffracted for a Bragg $\pi$-pulse with $\Omega = 2\pi \times 1.8$~kHz. Inset: zoom-in on the tail of $\tilde{n}_{\rm BEC} (k)$.}
 \end{figure}

For our $\Omega = 2\pi \times 1.8$~kHz, in Fig.~\ref{Fig1S} we indicate by the shadings the fractions of $\tilde{n}_{\rm BEC} (k)$ and $\tilde{n}_{\rm QD} (k)$ that are diffracted, for $\sqrt{na^3} = 0.04$ and our experimental values of $n$ and $L$.

 \begin{figure}[t!]
\centering
\includegraphics[width=\columnwidth]{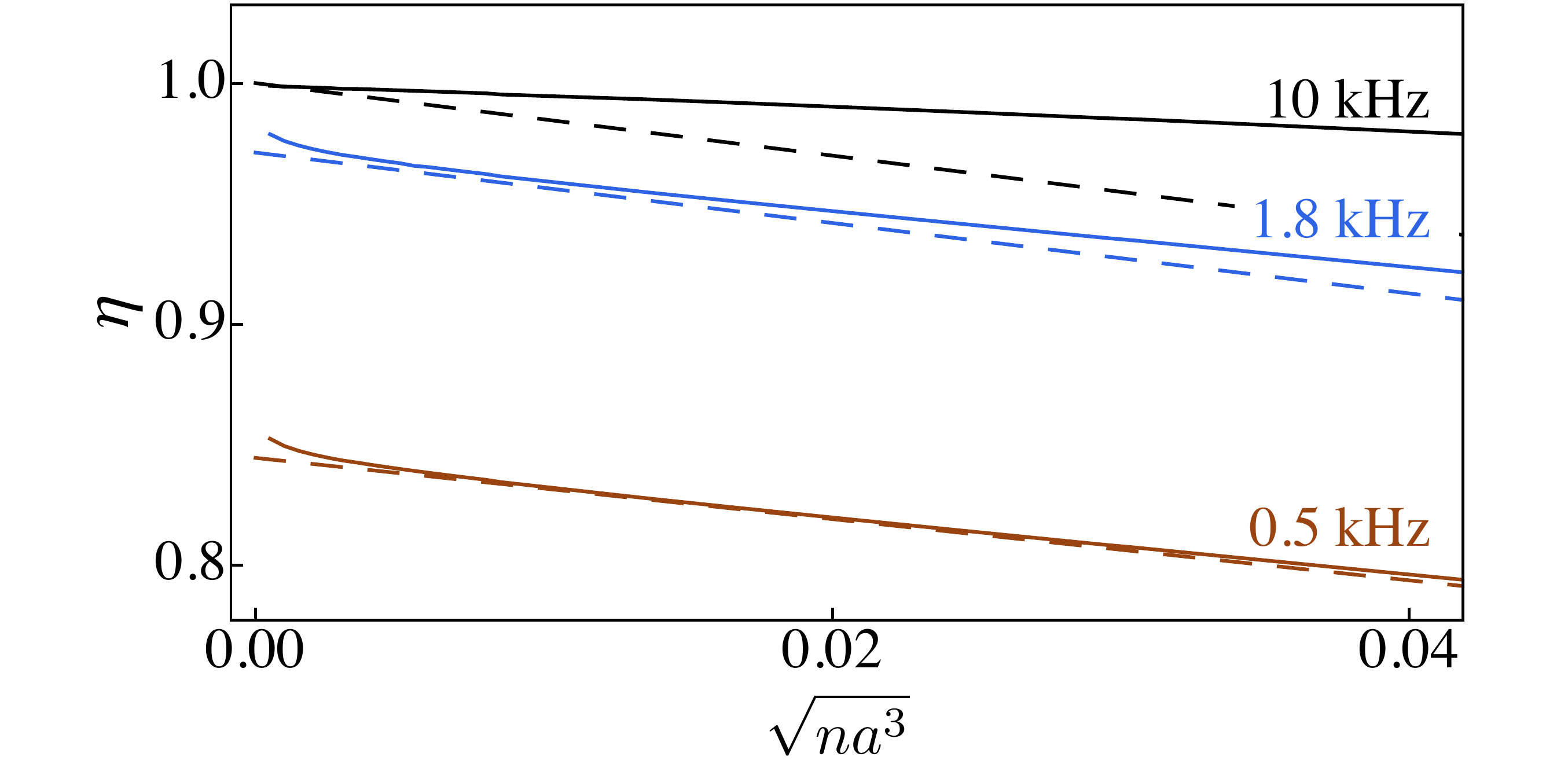}
\caption{\label{Fig2S}
Choice of $\Omega$.  We plot $\eta$ as a function of $\sqrt{na^3}$ for different values of $\Omega$ (solid lines); the dashed lines correspond to $\gamma =1.5$. In an ideal measurement, possible only for $L/\xi \rightarrow \infty$ (and $T=0$), one would observe $\eta(0) = 1$ and $\gamma=1.5$. In reality, for a non-infinite $L/\xi$, increasing $\Omega$ decreases the systematic effects on  $\eta(0)$, but increases the systematic effects on the observed $\gamma$. Our $\Omega = 2\pi \times 1.8$~kHz offers a good compromise for our system parameters.}
 \end{figure} 

Both the imperfect diffraction of the BEC and the partial diffraction of the QD introduce (opposite) systematic effects on the expected experimentally observed $\gamma$. The two effects partially cancel, but the latter is larger, and for any nonzero $\Omega$ the expected observed $\gamma$ is below $1.5$. 

In Fig.~\ref{Fig2S} we plot the simulated $\pi$-pulse diffracted fraction $\eta$ versus $\sqrt{na^3}$, for different values of $\Omega$; in all cases the dashed lines correspond to $\gamma =1.5$. For our $\Omega = 2\pi \times 1.8$~kHz, most of the BEC is diffracted ($\eta(0) > 97\%$), and the apparent reduction of $\gamma$ is only $\approx 20\%$ (comparable to our experimental errors). For much smaller $\Omega$ the expected $\gamma$ approaches the ideal theoretical value, but $\eta (0)$ drops significantly. Conversely, for much larger $\Omega$ essentially all of the BEC is diffracted ($\eta(0) \rightarrow 1$), but the expected key experimental signal, the interaction tuning of $\eta$, significantly diminishes, because most of the QD is diffracted as well. 

While these effects can be accounted for numerically, we use $\Omega$ for which the systematic effects on both $\eta(0)$ and $\gamma$ are small in the first place, so that numerical analysis is not essential for the interpretation of the experiments.

\subsection{Effects of nonzero temperature} 

To assess the effects of nonzero temperature, we start with the Bogoliubov dispersion relation
\begin{equation}
\varepsilon (k) = \sqrt{\frac{\hbar^2 k^2}{2m} \pc{\frac{\hbar^2 k^2}{2m} + 2 gn}} \, ,
\end{equation}
where $g = (4\pi \hbar^2 /m) a$, and assume that the depletion of the condensate is small. For any given $a$ one can compute, as a function of $T$, the energy density
\begin{equation}
E(T) - E(0)  = \frac{1}{(2\pi)^3} \int \varepsilon(k) f(k,T)  \, d\textbf{k} \, ,
\end{equation}
where $f(k,T) = [e^{\varepsilon(k)/(k_{\rm B} T)} -1]^{-1}$, the entropy density
\begin{equation}
S (T) = \int_0^T \frac{1}{T'} \frac{\partial E(T')}{\partial T'}  d T' \, , 
\end{equation}
and the thermal depletion
\begin{equation}
n_\text{ex} (T) =  \frac{1}{(2\pi)^3} \int p(k)  f(k, T)   \, d\textbf{k}  \,  ,
\label{eq:thermalDpl}
\end{equation}
where $p(k) = \left[ gn + \hbar^2 k^2/(2m) \right] / \varepsilon (k)$ is the particle content of a collective excitation mode~\cite{Pethick:2002}.

\begin{figure}[t!]
\centering
  \includegraphics[width=\columnwidth]{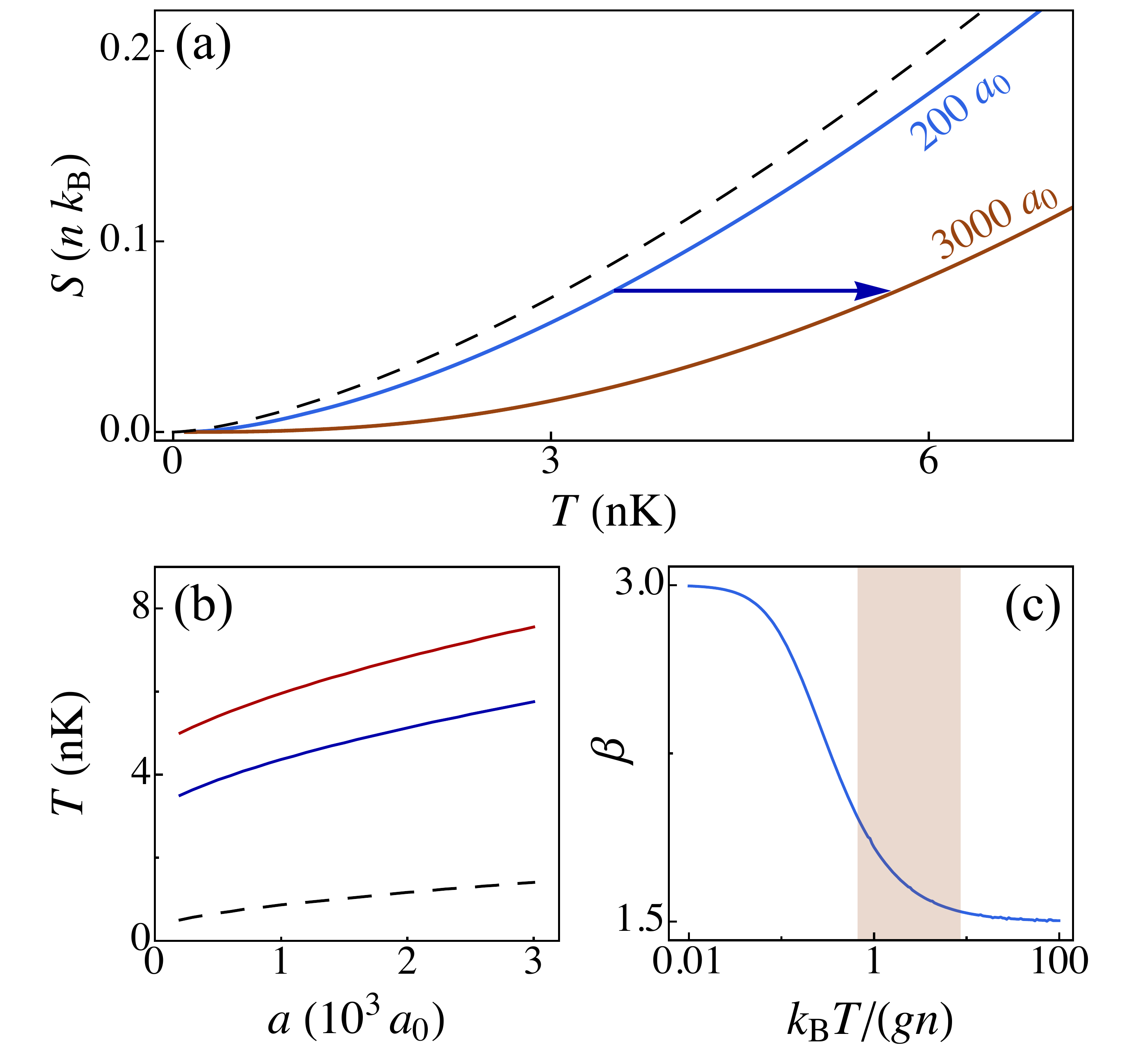}
\caption{\label{Fig3S}
Adiabatic heating. (a) Entropy $S(T)$ for $a=0$ (dashed line), $200~a_0$ (blue) and $3000~a_0$ (orange), for $n=3.5 \times 10^{11} \text{cm}^{-3}$. The arrow represents an adiabatic increase of $a$, which increases $T$. 
(b) Adiabatic evolution of $T$ with $a$, for initial $a=200~a_0$ and various initial temperatures.
(c) $\beta \equiv \partial \ln(S)/ \partial \ln(T)$ as a function of $k_{\rm B} T/(gn)$; see text.
}
 \end{figure} 
 
 In Fig.~\ref{Fig3S}(a), we plot $S(T)$ for $a=200~a_0$ and $3000~a_0$ (our experimental range); for reference we also show the $a=0$ result (dashed line). Qualitatively, for a larger $a$ the excitation energies $\varepsilon(k)$ are higher, so at the same $T$ there are less excitations and the entropy is lower. As indicated by the blue arrow, adiabatically increasing $a$ increases $T$. In Fig.~\ref{Fig3S}(b) we show how $T$ evolves for initial $a=200~a_0$ and various initial temperatures.

\begin{figure}[b!]
\centering
  \includegraphics[width=\columnwidth]{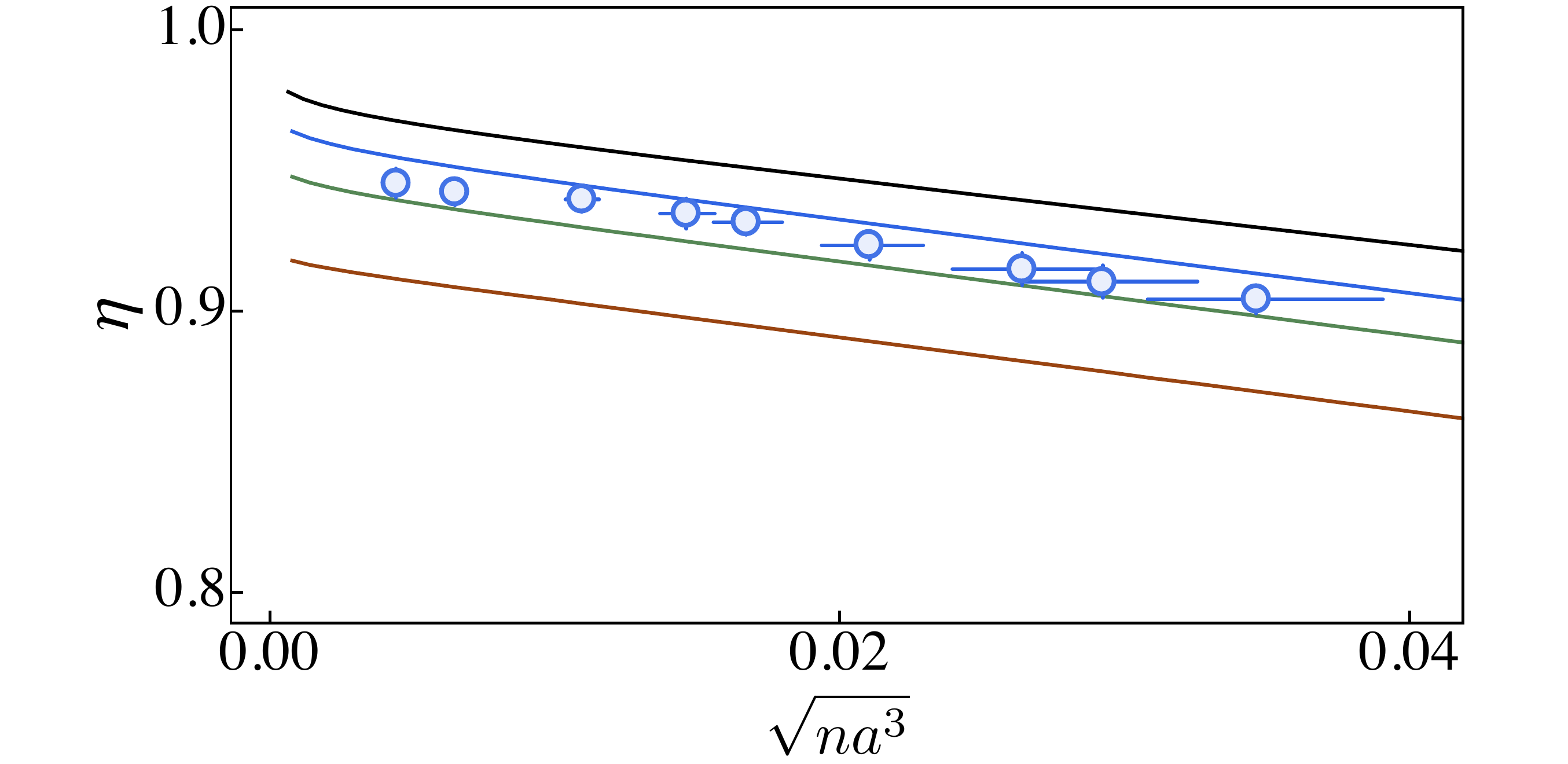}
\caption{\label{Fig4S}
Effects of the nonzero temperature on $\eta$. Here $\Omega =  2\pi \times 1.8$~kHz and the initial temperatures are, top to bottom, 0 (black),  3.5~nK (blue), 5~nK (green), and 7~nK (orange). Our experimental data (circles) are all modelled well by initial $T$ between 3.5 and 5~nK.}
 \end{figure} 

For intuition, it is instructive to consider the low-$T$, phonon-dominated regime, where the fractional adiabatic increase of $T$ is the largest, and one can analytically derive simple scalings. 
In this case $\varepsilon (k) = \hbar k c$, where $c = \sqrt{gn/m} \propto \sqrt{a}$ is the speed of sound, and $S \propto T^3/c^3$. Hence, for an adiabatic increase of $a$, we have $T\propto c \propto \sqrt{a}$. Intuitively, the energies of all the (already) thermally populated modes grow as $c \propto \sqrt{a}$, and hence so does $T$.  The occupation numbers of the phonon modes, $f(k,T) = [e^{\hbar k c/(k_{\rm B} T)} -1]^{-1}$, do not change, but their particle content does. We have $p(k) = gn/(\hbar k c) \propto c/k$ and $n_{\rm ex} \propto T^2/c$, so keeping $T/c$ constant increases the thermal depletion according to $n_{\rm ex} \propto T \propto \sqrt{a}$. 
For higher initial $T$ the fractional adiabatic increase of temperature is smaller because the energies of the higher-$k$ modes  ($k\gtrsim 1/\xi$) are less sensitive to the changes in $a$.

 In Fig.~\ref{Fig3S}(c) we plot $\beta \equiv \partial \ln(S)/ \partial \ln(T)$, which depends only on $k_{\rm B} T/(gn)$. For $k_{\rm B} T \ll gn$ and $k_{\rm B} T \gg gn$, respectively, $\beta$ approaches constant values of $3$ and $3/2$. This recovers two familiar results: for $k_{\rm B} T \ll gn$, the thermal excitations are phonons and $S \propto T^3$; in the opposite limit, assuming only particle-like excitations with a quadratic dispersion relation, $S \propto T^{3/2}$. 
Our experiments fall into the shaded region of the graph, so a numerical calculation is necessary.

In Fig.~\ref{Fig4S} we show numerical simulations of $\eta$ that include both quantum and thermal depletion, for  different initial temperatures (at $a=200~a_0$). As shown in the main paper, all our experimental data are modelled well by initial temperatures between 3.5 and 5~nK.

\end{document}